\documentclass[useAMS,usenatbib,usegraphicx]{mn2e}
\usepackage{times}
\usepackage{graphicx}
\usepackage{amssymb}
\usepackage{multirow}
\usepackage{color}
\input{epsf}

\newif\ifAMStwofonts
\AMStwofontstrue

\title[Formation of sub-millisecond pulsars and possibility of detection]
{Formation of Sub-millisecond Pulsars and Possibility of Detection}
\author[Y. J. Du, R. X. Xu, G. J. Qiao and J. L. Han]
{Y. J. Du,$^{1}$\thanks{E-mail: dyj@bao.ac.cn.}
R. X. Xu,$^{2}$~%\thanks{E-mail: r.x.xu@pku.edu.cn.}
G. J. Qiao,$^{2}$ and ~%\thanks{E-mail: gjn@pku.edu.cn,}
J. L. Han$^{1}$%\thanks{E-mail:hjl@bao.ac.cn.} \\
\\
$^{1}$National Astronomical Observatories, Chinese Academy of
Sciences, Jia-20, Datun Road, Chaoyang District, Beijing 100012, China \\
$^{2}$Department of Astronomy, Peking University, Beijing 100871,
China }

\pagerange{\pageref{firstpage}--\pageref{lastpage}}

\begin{document}

\maketitle

\label{firstpage}

\begin{abstract}
Pulsars have been recognized as normal neutron stars, but sometimes
argued as quark stars. {\it Sub-millisecond pulsars, if detected,
  would play an essential and important role in distinguishing quark
  stars from neutron stars.}
We focus on the formation of such sub-millisecond
pulsars in this paper. A new approach to form a sub-millisecond
pulsar (quark star) via accretion induced collapse (AIC) of a white
dwarf is investigated here. Under this AIC process, we found that:
(1) almost all the newborn quark stars could have an initial spin
period of $\sim 0.1$ ms; (2) the nascent quark stars (even with a
low mass) have sufficiently high spin-down luminosity and satisfy
the conditions for pair production and sparking process to be as
sub-millisecond radio pulsars; (3) in most cases, the timescales of
newborn quark stars in the phase of spin period  $< 1$ (or $<0.5$)
ms can be long enough to be detected.

As a comparison, an accretion spin-up process (for both neutron and
quark stars) is also investigated.  It is found that, quark stars
formed through AIC process can have shorter periods ($\leq$ 0.5\,ms);
while the periods of neutron stars formed in accretion spin-up process
must be longer than $0.5$ms. Thus if a pulsar with a period less than
$0.5$ ms can be identified in the future, it should be a quark star.

\end{abstract}


\section{Introduction}

Though it has been more than 40 years since the discovery of radio
pulsars, their real nature is still not yet clear because of the
uncertainty about cold matter at supranuclear density.
Both neutron matter and quark matter are two conjectured states for
such compact objects. The objects with the former are called 
neutron stars, and with the latter are quark stars.
It is an astrophysical challenge to observationally distinguish real
quark stars from neutron stars (see reviews by, e.g., Madsen 1999;
Glendenning 2000; Lattimer \& Prakash 2001; Kapoor \& Shukre 2001;
Weber 2005; Xu 2008).
The most obvious discrepancy could be the minimal spin period of these
two distinct objects. The minimal periods of these two kinds of
objects are related to their formation process. How fast a neutron
star or a quark star can rotate during the recycling process in low
mass X-ray binary (LMXB) has been considered by several authors
(Bulik, Gondek-Rosi$\rm \acute{n}$ska \& Klu$\rm \acute{z}$niak 1999;
Blaschke et al. 2002; Zdunik, Haensel \& Gourgoulhon 2002; Xu 2005;
Arras 2005).
Friedman, Parker \& Ipser (1984) have found that neutron stars with
the softest equation of state can rotate as fast as 0.4 ms. The smallest
spin period for neutron stars computed by Cook, Shapiro \& Teukolsky
(1994) is about 0.6 ms. Frieman \& Olinto (1989) have showed that the
maximum rotation rate of secularly stable quark stars may be less than
0.5 ms. Burderi \& D'Amico (1997) have discussed a possible
evolutionary scenario resulting in a sub-millisecond pulsar and
the possibility of detecting a sub-millisecond pulsar with a
fine-tuned pulsar-search survey. Gourgoulhon et al. (1999) have
investigated the maximally rotating configurations of quark stars and
showed that the minimal spin period was between 0.513 ms and 0.640
ms. Burderi et al. (1999) have predicted that there might exist a yet
undetected population of massive sub-millisecond neutron stars, and
the discovery of a sub-millisecond neutron star would imply a lower
limit for its mass of about 1.7 $M_\odot$. A detailed investigation
about spin-up of neutron stars to sub-millisecond period, including a
complete statistical analysis of the ratio with respect to normal
millisecond pulsars, was performed by Possenti et al. (1999). The
minimal recycled period was found to be 0.7 ms. Gondek-Rosinska et
al. (2001) have found that the shortest spin period is approximately
0.6 ms through the maximum orbital frequency of accreting quark
stars. Huang \& Wu (2003) have found that the initial periods of
pulsars are in the range of 0.6 $\sim$ 2.6 ms using the proper motion
data.  Zheng et al. (2006) have showed that hybrid stars instead of
neutron or quark stars may lead to sub-millisecond pulsars. Haensel,
Zdunik \& Bejger (2008) have discussed the compact stars' equation of
state (EOS) and the spin-up to sub-millisecond period, via mass
accretion from a disk in a low-mass X-ray binary.

There have been many observational attempts in searching
sub-millisecond pulsars. A possible discovery of a 0.5 millisecond
pulsar in Supernova 1987A is not held true in the follow-up
observations (Sasseen 1990; Percival et al. 1995). Bell
et al. (1995) reported on optical observation of the low mass binary
millisecond pulsar system PSR J0034-0534, and they used white dwarf
cooling models to speculate that, the limit magnitude of the J0034-0534's
companion suggested that this millisecond pulsar's initial spin period
was as short as 0.6 ms. As addressed by D'Amico \& Burderi (1999), in
particular the detection of a pulsar with a spin period well below 1
ms could put severe constraints on the neutron star structure and the
absolute ground state for the baryon matter in nature. 
They designed an experiment to find sub-millisecond pulsars with
Italian Northern Cross radio telescope near Bologna. Edwards, van
Straten \& Bailes (2001) have found none of sub-millisecond pulsars in
a search of 19 globular clusters using the Parks 64 m Radio telescope
at 660 MHz with a time resolution of 25.6 $\rm \mu s$. Han et
al. (2004) did not find any sub-millisecond pulsars from highly
polarized radio source of NVSS (NRAO VLA Sky Survey).  Kaaret et
al. (2007) have found oscillations at a frequency of 1122 Hz in an
X-ray burst from a transient source XTE J1739-285 which may contain
the fastest rotating neutron star so far. Significant difficulties do
exist in current radio surveys for binary sub-millisecond pulsars due
to strong Doppler modulation and computational limitations (Burderi et
al. 2001).

How do sub-millisecond pulsars form? This is still an open question
which we will explore in this paper.
Previously, discussions were concentrated on the formation of neutron
stars or quark stars spun up via accretion in binaries.  We have
considered a new approach to create a sub-millisecond pulsar (quark
star) with super-Keplerian spin via accretion induced collapse (AIC)
of a massive white dwarf (WD).
%This scenario has never been considered in details before.
The initial spin of the newborn quark star could be super-Keplerian,
and it can have a long lifetime in sub-millisecond phase and produce
enough strong radio emission to be detected.

%arrangement:
In $\S2$, we discuss low mass quark stars formed from AIC of WDs,
which can have an minimal initial period of sub-millisecond.
In $\S3$, the radiation parameters and the conditions for pair
production are estimated in order to investigate whether the
AIC-induced quark stars could be pulsars or not. The lifetimes of
sub-millisecond pulsars are also estimated and the possibility of
detection is discussed. The spin-down evolution diagrams of a newborn
quark star and neutron star are also plotted.
In $\S 4$, as a comparison, the sub-millisecond pulsars formed through
accretion acceleration (spin-up) in binary systems are also considered.
In $\S 5$, conclusions and discussions are presented.

\section{Sub-millisecond quark stars formed through
AIC of white dwarfs}

Neutron star's formation from AIC of a massive white dwarf is widely
discussed by many authors (Nomoto et al. 1979; Nomoto \& Kondo 1991;
van Paradijs et al. 1997; Fryer et al. 1999; Bravo \&
Garc{\'{\i}}a-Senz 1999; Dessart et al. 2006). Recently, it is pointed
out that Galactic core-collapse supernova rate cannot sustain all the
separate neutron star populations (Keane \& Kramer 2008), which
implies other mechanisms for forming neutron stars. AIC of a massive
WD can be an important mechanism for pulsar formation, even for
isolated pulsars if the binary systems are destroyed due to strong
kicks. We now discuss the possibility for a low mass quark star formed
from AIC of a WD. In a binary system, when the WD has accreted enough
matter from the companion so that its mass reaches the Chandrasekhar
limit, the process of electron capture may induce gravitational
collapse. The detonation waves burn nuclear matter into strange quark
matter which spread out from the inner core of the WD (Lugones,
Benvenuto \& Vucetich 1994). A boundary of strange quark matter and
nuclear matter will be found at the radius where the detonation waves
stop when nuclear matter density drops below a critical value. A
similar process was also discussed and calculated by Chen, Yu \& Xu
(2007). The size of the inner collapsed core may depend on the chemical
composition and accretion history of the WD (Nomoto \& Kondo
1991). Consequently, quark stars with different masses could be formed.

\begin{figure}
\vspace{8.2cm}
\includegraphics{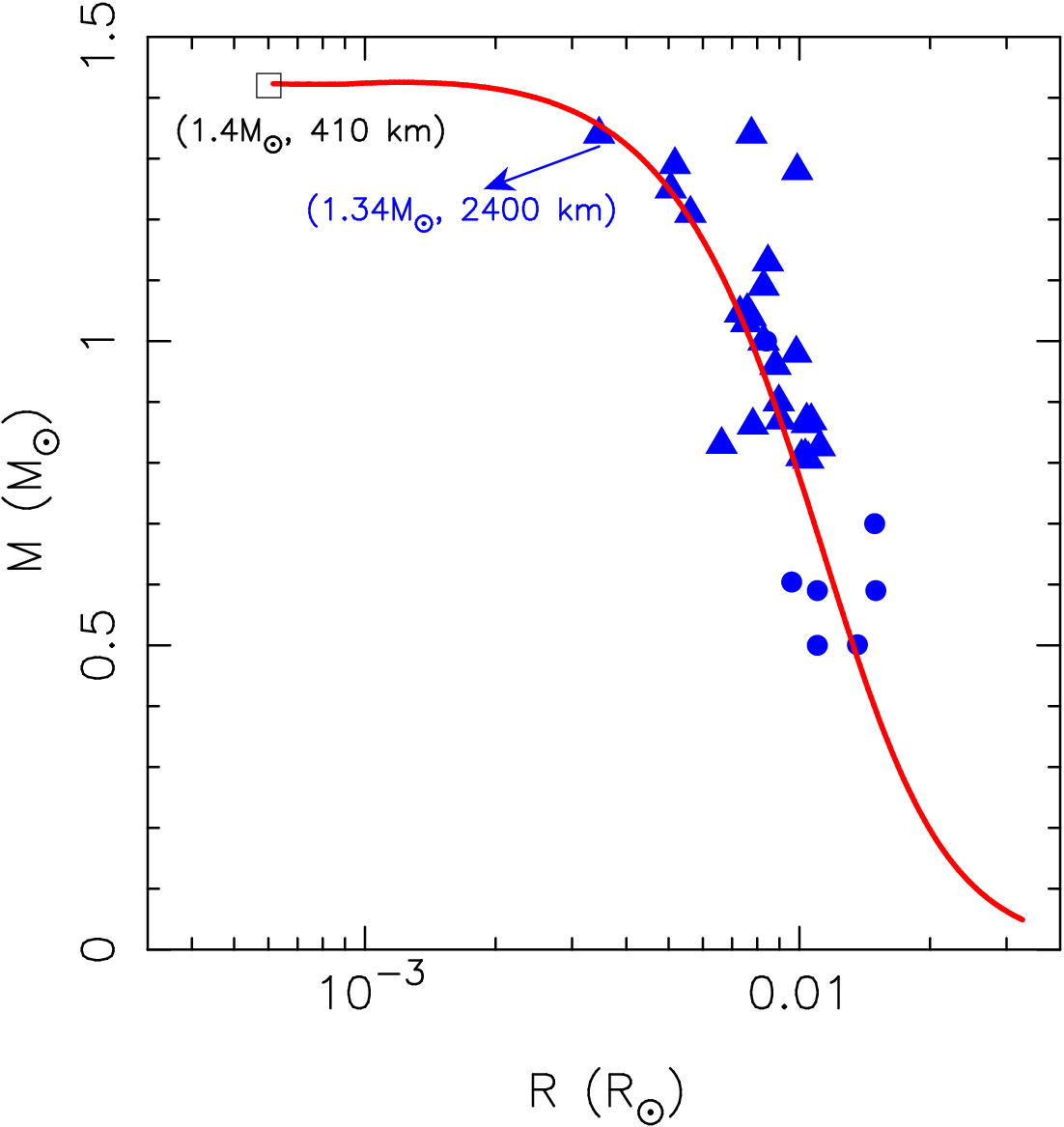}
\caption {The relation of mass and radius of WDs. The red line is
theoretical line.
%calculated which we adopted the central density
%$\rho_{\rm c}$ from $10^4$ to $10^{11}$ g\,cm$^{-3}$.
The blue triangles and circles are observed WDs' data which were
taken from Table 1 of Nale{\.z}yty \& Madej (2004) and Table 3 \& 5
of Provencal et al. (1998), respectively. Among these data, the WD RE
J0317-853 is the most massive WD, whose mass and radius are 1.34$M_\odot$
and 2400 km respectively. The square is the point of a WD with the 
Chandrasekhar mass limit.} \label{fig1}
\end{figure}
%
%On one hand,
Both rigidly and differentially rotating WDs are taken into
account. As a first step, we assume that both the collapsed WD and the
newborn quark star have rigidly rotating configurations for
simplicity. The WDs, progenitors of these quark stars, could have a
uniformly rotating configuration due to the effects of
crystallization, as well as an increase of central density may lead to
catastrophic evolution (supernova) (Koester 1974). With these
assumptions, a model of sub-millisecond pulsars' formation is given
below.
The initial spin period of AIC-produced quark stars can be estimated
as follows. We assume that the mass ($M_\star$) of the nascent quark star
ranges from $10^{-3}M_\odot$ to $1M_\odot$, and the white dwarf
rotates rigidly at the Kepler period ($P_{\rm K}$) just before
collapsing. The quark star's rest mass ($M_\star$) is approximately
equal to the mass ($m_{\rm core}$) of the inner collapsed core of the
white dwarf. If the angular momentum is conserved during AIC, the
newborn quark star can rotate at a much shorter period, $P_{\rm q}$,
then
\begin{equation}
I_{\rm core}\frac{2\pi}{P_K} = I_{\rm q}\frac{2\pi}{P_{\rm q}}.
\end{equation}%
%\label{pss}
This is to say,
\begin{equation}
P_{\rm q} = \frac{I_{\rm q}}{I_{\rm core}}P_{\rm K},
\end{equation}
where $I_{\rm q}$ is the quark star's moment of inertia, and $I_{\rm
core}$ is the moment of inertia of WD's inner collapsed core, which
can be well approximated by
\begin{equation}
I_{\rm core} \simeq \frac{2}{5}M_{\rm core} R_{\rm core}^2.
\end{equation}
%
%
%This implies that a quark star could be uniformly rotating almost
%initially.
The mass and radius of a low mass $(M_\star \leqslant 1M_\odot)$
quark star could be approximately related by $M_\star = (4/3)\pi
(4\beta)R^3$ (Alcock, Farhi \& Olinto 1986) in the bag model. We
have an approximate formula for the fast rotating quark star's moment
of inertia
% which does not affect the astrophysical discussion on
%this problem, i.e.,

\begin{eqnarray}
%I_{\rm q} &=& \int_0 ^{R} \frac{4\beta 4\pi r^4} {\sqrt{1-\frac{4\pi^2
%r^2}{c^2 P_{\rm q}^2}}} dr = \frac{\beta}{16\pi^4}[3c^5 P_{\rm q}^5
%\arcsin({\frac{2\pi R}{cP_{\rm q}}}) \nonumber\\ &-& 2\pi c^2 P_{\rm
%q}^2 R\sqrt{1-\frac{4\pi^2 R^2}{c^2 P_{\rm q}^2}}(3c^2P_{\rm
%q}^2+8\pi^2 R^5)]
I_{\rm q} &=& 2\int_0 ^{R}  d z \int_0 ^{\sqrt{R^2-z^2}}
\frac{4\beta 2\pi x^3} {\sqrt{1-\frac{4\pi^2 x^2}{c^2 P_{\rm
q}^2}}} d x \nonumber\\ &=& \frac{\beta c P_{\rm q}}{16 \pi^4}[6\pi
c^3 P_{\rm q}^3 R- 8\pi^3 c P_{\rm q}R^3 + (16\pi^4 R^4 \nonumber\\
&+&8\pi^2 c^2 P_{\rm q}^2 R^2 -3c^4 P_{\rm q}^4) \ln
\frac{1+\frac{2\pi R}{c P_{\rm q}}}{\sqrt{1-\frac{4\pi^2 R^2}{c^2
P_{\rm q}^2}}}],
\end{eqnarray}
where z-axis is spin axis; x is integral variable of each disc
perpendicular to the spin axis; $c$ is the speed of light; $R$ is the
quark star's radius; the bag constant $\beta$ of quark stars is
(60--110) MeV fm$^{-3}$, i.e.  (1.07--1.96) $\times 10^{14}$ g
cm$^{-3}$, $\beta_{14}$ in units of $10^{14}~\rm g ~cm^{-3}$.

For WD, we made a code, using both non-relativistic hydrostatic
equilibrium equation
\begin{equation}
\frac{dp}{dr} = - \frac{Gm(r)\rho(r)}{r^2},
\end{equation}
and general equation of state (EOS) for a completely degenerate fermi
gas
\begin{eqnarray}
p &=& \frac{1}{3\pi^2\hbar^3} \int_0 ^{p_F} \frac{c^2p^4}{\sqrt{c^2p^2
+m^2c^4}} d p \nonumber\\
&=& 1.42\times10^{25}\phi(x) ~\rm dyn\; cm^{-2},
\end{eqnarray}
where $x\equiv p_F/mc$, $\lambda_e=\hbar/(mc)$ the electron's
Compton wavelength, $P_{\rm F}$ the fermi momentum,
\begin{eqnarray}
\phi(x) & = & (8\pi^2)^{-1}\{x(1+x^2)^{1/2}(2x^2/3-1) \nonumber\\
&&+\ln[x+(1+x^2)^{1/2}]\}\nonumber
\end{eqnarray}
to calculate the mass ($m_{\rm core}$) and moment of inertia ($I_{\rm
core}$) of the collapsed core of a massive WD, where $p$,
$\rho$, $G$, $\hbar$ are pressure, mass density, the gravitational
constant and the Planck constant, respectively.

Using Eqs. (5) and (6), one can make numerical calculation to get the
WD's theoretical relation of mass and radius (the red line in Figure
1). For comparison, one can see a figure on line \footnote{{\tt
    http://cococubed.asu.edu/code{\_}pages/coldwd.shtml}}. Before
collapsing, the mass of a WD is close to the Chandrasekhar mass limit,
as high as $M_{\rm WD}=1.4M_\odot$ (Shu 1982), the corresponding
radius is much smaller, such as $R_{\rm WD}=410$\,km.
We could numerically obtain the initial period, $P_{\rm q}$, of
nascent quark stars with different mass via Eq. (2), and find that
almost all the values of $P_{\rm q}$ are around $\sim 0.1$ ms (See
Table 1) if the WD rotates rigidly at an almost Kepler period due to
accretion (or spin-up) in a binary just before collapsing.
The newborn quark stars' surface spin velocities are well
above the Kepler velocities, we regard this as ``the super-Keplerian
case''.  

WDs may be rotating differentially. The detailed calculations are
given in Appendix A. Therefore, as a follow-up second step, we also
use Eqs. (2), (5), (6) and (A2) to calculate the initial spin period
of the nascent quark stars in the differentially rotating WD model,
taking the free parameter $a=0.5$. The results are also shown in Table 1.

A newborn quark star could certainly rotate differentially, and may
be relaxed to become a rigidly rotating configuration finally.
However, the timescale of the relaxation depends on the viscosity
and the state of cold quark matter ~\citep{xu2009}. Nevertheless,
the newborn quark star's relaxation (from differentially
rotating configuration to rigidly rotating configuration) may be due
to fast solidification after birth. A calculation shows that the
solidification timescale is only $10^3-10^6$ s~\citep{xl2009}.
%which is calculated by a model
% if quark stars could be solidified.
Therefore, the relaxation timescale could be much shorter than the
lifetime of pulsars within sub-millisecond periods (See the following
section 3.3).

The WD RE J0317-853 has the highest observed mass (1.34 $M_\odot$
close to the Chandrasekhar limit) with radius of 2400 km (Nale{\.z}yty
\& Madej 2004). If a WD like RE J0317-853 could be in a binary and
accreted enough materials to the Chandrasekhar limit, then it may
collapse. Therefore, under this assumption, we also calculated the
initial spin periods $\widehat{P}_{\rm q}$ and $\widehat{P}_{\rm dif}$
of a nascent quark star. The calculated results are listed in Table
1. It is found that, even if a WD has a larger radius such as 2400 km,
it can also collapse to a sub-millisecond quark star for either
rigidly or differentially rotating WD models.
%the resultant initial periods
%of quark stars can also be $< 1$ millisecond, being similar to the
%values obtained in the rigidly rotating WD model. Additionally,
In the differentially rotating WD model, it tends to give a rigidly
rotating configurations in the limit of large values of $a$, $P_{\rm
dif}$ increases as the parameter $a$ increases. The conclusions from
the rigid rotation model are valid even if differential rotation is
included.
%
%Actually the stellar mass should be higher than $1.34 M_\odot$ (the
%stellar radius thus becomes smaller) before real collapsing, and the
%conclusion still keeps.}
%

Can a quark star survive even if it rotates at such a high frequency
($\sim 10^4$ Hz)? Will it be torn apart by the centrifugal force?
There are quite distinguishing characteristics between neutron stars
and quark stars.
A low mass quark star is possible to spin at a super-Keplerian
frequency because it is self-bound by strong interaction.
On one hand, as noted by Qiu \& Xu (2006), astrophysical quark
matter splitting could be color-charged if color confinement cannot
be held exactly because of causality. On the other hand, however,
rapidly spinning quark matter could hardly split if color confinement
is held exactly.
In addition, the recently discovered nature of strongly coupled quark
gluon plasma (sQGP) as realized at Relativistic Heavy Ion Collider
(RHIC) experiment (e.g., Shuryak 2006) may also prevent a
super-Keplerian quark star to split.

The short spin period above is not surprising, and could be verified
for a simplified special case, if both quark star's density (=4$\beta$) and
white dwarf's density ($=\rho_c$) are uniform. Using Eq. (2) and the
mass-radius relation, we can find the initial period of the quark star
to be $P_{\rm q}=(\rho_c/4\beta)^{2/3} P_{\rm WD}\sim 4\times
10^{-3}(\rho_{11}/\beta_{14})^{2/3}P_{\rm WD}$ (with $P_{\rm WD}$ the
spin period of white dwarf, $\rho_{11}=\rho_c/10^{11}$g cm$^{3}$,
$\beta_{14}=\beta /10^{14} \rm g$ cm$^3$), which depends only on the
densities of WD and quark star.

If the WD has not been spun up fully to the Kepler period, i.e.,
the WD rotates at a sub-Keplerian period (e.g., several times of
$P_{\rm K}$) before AIC, can the initial period of a newborn quark
star formed from such a WD be of sub-millisecond?  We investigated the
case of a massive WD (1.4$M_\odot$, 410 km) rotating at a period
$P_{\rm WD}=5P_{\rm K}\sim 600$\, ms. The initial spin period of quark
stars with different mass are as follows: $\hat{P}_{\rm q}\sim 0.11$
\,ms for a quark star with mass of $0.001M_\odot$; $\sim 0.24$\,ms for
$0.01M_\odot$; $0.35$\,ms for $0.1M_\odot$ and $ \sim 0.36$\,ms for $1
M_\odot$.
%It is evident that the initial
%period of quark star could be of sub-millisecond even if the WD
%rotates at five times of the Keplerian period before AIC.
%
%It is evident that $P_{\rm q}\sim 0.1$ ms if $P_{\rm WD}=P_{\rm
%K}$.The initial period of quark star could be of sub-millisecond even if
%As addressed in Eq.(2), $P_{\rm q}\propto P_{\rm WD}$, we may have
%$P_{\rm q} \simeq 0.3$ ms when $P_{\rm WD}\simeq 5P_{\rm K}=600$ms.
%
The spin-down feature of such a newborn quark star depends on its
gravitational wave radiation and magnetodipole radiation (see details
in $\S 3$). 
%We use the $P_{\rm q}$ (Column 3 in Table 1) of the
%rigidly rotating WD model to do the related calculations in the
%follow-up sections.}

\begin{table*}
\centering
\begin{minipage}{140mm}
\caption
{The minimal initial period ($P_{\rm q}$) and lifetimes ($\tau$)
due to GW and EM radiation in the phase of sub-millisecond period for
quark stars with different masses in the super-Keplerian case. $P_{\rm
  q}$ and $P_{\rm dif}$ are calculated via angular momentum
conservation using rigidly and differentially rotating WD model with
central density of $10^{11} \rm g\,cm^{-3} $. $\widehat{P}_{\rm q}$
and $\widehat{P}_{\rm dif}$ are similarly calculated but using a WD
like RE J0317-853 with mass of 1.4$M_\odot$ and radius of 2400 km.
$\tau_1$ is quark stars' lifetime in the phase of $<1$\,ms, while
$\tau_2$ is the timescale in the phase of $<0.5$\,ms.
$P_{\rm q}$ is also used in Table 2 \& 3 and Figure 2.  $\beta$ is the
bag constant, $\varepsilon_e$ is the gravitational ellipticity.\label{tbl-1}}
\begin{tabular}{lccccc}
\hline
Mass & Radius\, (km) & $P_{\rm q}$(ms) & $P_{\rm dif}$(ms)
& $\widehat{P}_{\rm q}$(ms) & $\widehat{P}_{\rm dif}$(ms)
\\
$(M_{\odot})$ & $\beta=60\rm~MeV~fm^{-3}$ & $\beta=60\rm~MeV~fm^{-3}$
& $a=0.5$ & $\beta=60\rm~MeV~fm^{-3}$ & $a=0.5$ \\
%\multirow{Mass} & \multirow{Radius\, (km)} &
%\multicolumn{2}{c}{$\rho_{\rm c}=10^{11}\,\rm g\, cm^{-3}$} &
%\multicolumn{2}{c}{$\rho_{\rm c}=10^{10}\,\rm g\, cm^{-3}$} &$\tau$(yr) \\
%\cline{3-4} \cline{5-6}
% & & $P_{\rm q}$(ms) & $P_{\rm dif}$(ms)
%& $\widehat{P}_{\rm q}$(ms) & $\widehat{P}_{\rm dif}$(ms) &
%\multicolumn{2}{|c|}{}  \\
%$(M_{\odot})$ & $\beta=60\rm~MeV~fm^{-3}$ & $\beta=60\rm~MeV~fm^{-3}$
%& $a=0.5$ & $\beta=60\rm~MeV~fm^{-3}$ & $a=0.5$ & $\varepsilon_e=10^{-6}$ \\
%\hline
% & & \multicolumn{2}{c}{$\rho_{\rm c}=10^{11}\,\rm g\, cm^{-3}$ }
%&\multicolumn{2}{c}{$\rho_{\rm c}=10^{10}\,\rm g\, cm^{-3}$} & \\
\hline
0.001 & 1.04    & 0.0699 & 0.0261 & 0.0481 & 0.0252  \\
0.01  & 2.24    & 0.0751 & 0.0472 & 0.0512 & 0.0470  \\
0.1   & 4.81    & 0.104  & 0.101  & 0.102  & 0.101   \\
1     & 10.37   & 0.221  & 0.218  & 0.218  & 0.218  \\
\hline
\end{tabular}

\begin{tabular}{|l|c||c|c|c|c|}
%\centering
\hline
\multirow{2}{*}{Mass($M_{\odot}$)} & \multirow{2}{*}{$P_{\rm q}$(ms)} & 
\multicolumn{2}{|c|}{$\tau_1$(yr)} 
& \multicolumn{2}{|c|}{$\tau_2$(yr)}\\
%\cline{3-6}
& & $\varepsilon_e=10^{-6}$ & $\varepsilon_e=10^{-9}$  
& $\varepsilon_e=10^{-6}$ & $\varepsilon_e=10^{-9}$\\
\hline
0.001  & 0.0699  & $3.4\times10^7$ & $4.5\times10^{10}$
& $2.1\times10^{6}$ & $1.1\times10^{10}$   \\
0.01   & 0.0751  & $7.3\times10^5$ & $2.0\times10^{10}$
& $4.5\times10^{4}$  & $4.3\times10^{9}$      \\
0.1    & 0.104   & $1.6\times10^4$ & $5.4\times10^9$
& $9.0\times10^{2}$ & $6.5\times10^{8}$   \\
1      & 0.221   & $3.4\times10^2$ & $3.1\times10^{8}$
& 2.2$\times10^{1}$     & $2.0\times10^{7}$   \\
\hline
\end{tabular}
\end{minipage}
\end{table*}

\section{Radiation of sub-millisecond quark stars with low masses}

The mass of most sub-millisecond quark stars formed from WD's AIC is
so low, can the quark star produce radiation luminous enough to be
observed like millisecond pulsars? This is related to two aspects.
First of all, is the rotational energy loss rate high enough to power
the electromagnetic radiation as normal neutron stars?  Secondly, is
the potential drop in the inner gap high enough for pair
production and sparking to take place in the inner gap? These are
necessary conditions for radio emission of pulsars.

\subsection{The spin-down power of sub-millisecond pulsars}

Normal radio pulsars are rotation-powered, and the radiation energy is
coming from the rotational energy loss. Here we neglect gravitational
wave radiation first, then the rate $ \dot{E}_{\rm rot}$, is
\begin{equation}
\dot{E}_{\rm rot}= \frac{8 \pi^4 R^{6} B^2
P^{-4}}{3 c^3}.
\end{equation}
Comparing the rotational energy loss rate ($\dot{E}_{\rm rot,q}$) of
quark stars with normal neutron stars' ($\dot{E}_{\rm rot,NS}$), one
can have
\begin{equation}
\dot{E}_{\rm rot,q}/\dot{E}_{\rm rot,NS}= \frac{ R^{6}_{\rm q} B_{\rm q}^{2}
P^{-4}_{\rm q}}{R^{6}_{\rm NS} B_{\rm NS}^{2} P^{-4}_{\rm NS}}.
\end{equation}
If we take normal parameters, such as the surface magnetic field
of polar cap $B_{\rm q}=10^8$~G, $B_{\rm NS}=10^{12}$~G, the
rotational period $P_{\rm q}=0.1$~ms and $P_{\rm NS}=1$~s, the result
is $\dot{E}_{\rm rot,q}/\dot{E}_{\rm rot,NS}=10^2$ even for a quark
star with a mass of $0.001M_{\odot}$. This means that the quark stars
have enough rotational energy to radiate, hundred times than normal
pulsars, even if the mass is so low.

\begin{table*}
 \centering
 \begin{minipage}{140mm}
 \caption{Gap parameters estimated for sub-millisecond quark
 stars. $\dot{E}_ {\rm rot}$ is the spin-down luminosity; $h_{\rm CR}$
 is the curvature radiation (CR) gap height; $\Delta V_{\rm CR}$ is
 the potential drop of CR gap; $h_{\rm res}$ is the height of resonant
 ICS gap; $\Delta V_{\rm res}$ is the potential drop of resonant ICS
 gap; $h_{\rm th}$ is the thermal ICS gap height; $\Delta V_{\rm th}$
 is the potential drop of thermal ICS gap. \label{tbl-2}}
\begin{tabular}{cccccccc}
\hline $M (M_\odot)$ & $\dot{E}_{\rm rot}$(erg \,s$^{-1}$) &
$h_{\rm CR}$(cm) & $\Delta V_{\rm CR}$(V) & $h_{\rm res}$(cm)
& $\Delta V_{\rm res}$(V) & $h_{\rm th}$(cm) & $\Delta V_{\rm th}$(V) \\
\hline
0.001 & $4.99\times10^{36}$ & $1.16\times 10^4$ & $2.84\times 10^{10}$
& $3.13\times10^5$ & $2.05\times10^{13}$ & $1.18\times10^3$ &
$2.93\times10^8$ \\
0.01 & $3.75\times10^{38}$ & $1.34\times10^4$ & $3.76\times10^{10}$ &
$3.64\times10^{5}$ & $2.78\times10^{13}$ & $1.31\times10^3$ &
$3.62\times10^8$ \\
0.1 & $1.02\times10^{40}$ & $1.72\times10^{4}$ & $6.19\times10^{10}$ &
$4.61\times10^5$ & $4.46\times10^{13}$ & $1.62\times10^3$ &
$5.47\times10^8$ \\
1 & $5.00\times10^{40}$ & $2.65\times10^4$ & $1.47\times10^{11}$ &
$6.74\times10^5$ & $9.52\times10^{13}$ & $2.36\times10^3$ &
$1.16\times10^9$ \\
\hline
\end{tabular}
\end{minipage}
\end{table*}

\subsection{Particle acceleration for sub-millisecond pulsars}

In most radio emission models of pulsars, such as RS model (Ruderman
\& Sutherland 1975, hereafter RS75), inverse Compton scattering (ICS)
model (Qiao \& Lin 1998), the multi-ring sparking model (Gil \& Sendyk
2000), the annular gap model (Qiao et al. 2004) and so on, the
potential drop in the inner gap must be high enough so that the pair
production condition can be satisfied.

In the inner vacuum gap model, there is strong electric field parallel
to the magnetic field lines due to the homopolar generator effect. The
particles produced through $\gamma -B$ process in the gap can be
accelerated to ultra-relativistic energy (i.e., the lorentz factor can
be $10^6$ for normal pulsars). The potential across the gap is (RS75)

\begin{equation}
\bigtriangleup V=\frac{\Omega B}{c}h^2,
\end{equation}
where $\Omega$ is the angular frequency of the pulsar; $h$ is the gap
height; $B$ and $c$ represent the magnetic field at the surface of the
neutron star and the speed of light, respectively. As $h$ increases
and approaches $r_p$, the potential drop along a field line
traversing the gap can not be expressed by Eq. (9) above. In this
case the potential can reach a maximum value

\begin{equation}
\bigtriangleup V_{\max }=\frac{\Omega B}{2c}r_p^2,
\end{equation}
where $r_p$ is the radius of the polar cap.

Let us make an estimate about the quark star's potential drop
$\bigtriangleup V_{\rm q}$ in the polar gap region

\begin{equation}
\bigtriangleup V_{\rm q}=\frac {\Omega B_{\rm q}}{2 c}r_{\rm p,q}^{2},
\end{equation}
where $\Omega=2 \pi/P_{\rm q}$, $r_{\rm p,q}=R_{\rm q}(2 \pi R_{\rm
q}/{c P_{\rm q}})^{1/2}$. For normal neutron stars, $\bigtriangleup
V$ can be obtained by just changing the subscript q to NS. Thus 

\begin{equation}
\frac{\bigtriangleup V_{\rm q}}{\bigtriangleup V_{\rm NS}}=\frac {
B_{\rm q} R_{\rm q}^{3} P_{\rm q}^{-2} }{B_{\rm NS} R_{\rm NS}^{3}
P_{\rm NS}^{-2}}.
\end{equation}

As one can take $R_{\rm q}=1$~km for a quark star with the mass of
$0.001M_\odot$, $B_{\rm q}=10^8$~G, $B_{\rm NS}=10^{12}$~G, $R_{\rm
NS}=10$~km, $P_{\rm q}=0.1$~ms and $P_{\rm NS}=1$~s, we find that
$\bigtriangleup V_{\rm q}/\bigtriangleup V_{\rm NS}=10$. This means
that the quark stars can have enough potential drops in the polar cap
regions.

In the inner gap model, $\gamma-B$ process plays a very important
role, two conditions should be satisfied at the same time for pair
production: (1) to produce high energy $\gamma$-ray photons, a strong
enough potential drop should be reached; (2) for pair production, the
energy component of $\gamma$-ray photons perpendicular to the magnetic
field must satisfy $E_{\gamma,\perp} \geq 2 m_e c^2$ (Zhang \& Qiao
1998).

Particles produced in the gap can be accelerated by the electric
field in the gap and the Lorentz factor of the particles can be
written as
\begin{equation}
\gamma =\frac{e\bigtriangleup V}{m_ec^2},
\end{equation}
where $\gamma$ is the Lorentz factor of the particles accelerated by
the potential $\bigtriangleup V$, $m_e$ the mass of an electron or
positron, $e$ the charge of an electron.

%{\bf Pair production in the gap}.
In $\gamma-B$ process, the conditions for pair production are that the
mean free path of $\gamma$-ray photon in strong magnetic field is
equal to the gap heights, $l\approx h$. The mean free path of
$\gamma$-ray photon is given by (Erber 1966)
\begin{equation}
l=\frac{4.4}{e^2/\hbar c}\frac \hbar {m_ec}\frac{B_c}{B_{\perp
}}\exp (\frac 4{3\chi }),
\end{equation}
where $B_{\rm c}=4.414\times 10^{13}$~G is the critical magnetic field,
$\hbar$ the Planck's constant,
\begin{equation}
\chi =\frac{E_\gamma }{2m_ec^2}\sin \theta \frac
B{B_{\rm c}}=\frac{E_\gamma }{2m_ec^2}\frac{B_{\perp }}{B_{\rm c}},
\end{equation}
and $B_\perp$ is the magnetic field perpendicular to the moving
direction of $\gamma$ photons, which can be expressed as (RS75)
\begin{equation}
B_{\perp }\approx \frac h\rho B\approx \frac l\rho B.
\end{equation}
Here $l\approx h$ is the condition for sparks (pair production) to
take place. $\rho$ is curvature radius of the magnetic field
lines. For dipole magnetic configuration, it is (Zhang et al. 1997a)
\begin{equation}
\rho \approx \frac 43(\lambda Rc/\Omega )^{1/2}.  \label{dipole}
\end{equation}
where $\lambda$ is a parameter to show the field
lines, $\lambda=1$ corresponding to the last opening field line.
Gamma-ray energy from the curvature radiation process can be
written as 
\begin{equation}
E_{\gamma,cr}= \hbar \frac {3 \gamma^3 c }{2 \rho}. \label{dipole}
\end{equation}
We estimated the gap heights based on Zhang, Qiao \& Han (1997b), i.e.
\begin{eqnarray}
h_{\rm CR}\simeq 10^6
P^{3/7}B_{8}^{-4/7}\rho_6^{2/7} \rm cm.
\end{eqnarray}

When the relevant parameters used are $B=10^8$~G , $P=P_{\rm q}$ and
assuming a dipole magnetic configuration, for any mass quark stars,
one can estimate the gap height from curvature radiation (CR)
$h_{cr}\approx 10^4 \rm ~cm =100 ~\rm m$. This means that even if
without multipolar magnetic field assumption, the quark star can
still work well for the CR pair production.

There are three gap modes for pair production, i.e. resonant ICS mode,
thermal-peak ICS mode and CR mode (Zhang et al. 1997a). Each mode has
relevant gap parameters including gap potential drop $\Delta V$ and
the mean free path $l$ of $\gamma-B$ process. For normal neutron
stars, one needs the assumption of a multipolar magnetic field,
$\rho=10^6$~cm, as RS75; but for $0.1$~ms low mass quark stars, the
dipole curvature radius is about $10^6$~cm. We estimated gap
heights and other parameters based on the work of Zhang, Qiao \& Han
(1997b), as shown in Table 2.

One can see from Table 2 that when the high energy gamma-ray photons
come from resonant photon production, the height of the gap is
larger. For the thermal-peak ICS mode, it is one order of magnitude
lower than the CR mode, and two order of magnitude lower than resonant
ICS mode. This means that in most cases, the thermal-peak ICS induced
pair production is dominated in the gap.

The newborn sub-millisecond quark stars have enough spin-down
luminosities and gap potential drops (see Table 2), so that they may
emit radio or $\gamma$-ray photons with sufficient luminosities, which
can be detected by new facilities, e.g., FAST and Fermi (formerly
GLAST).

\subsection{Lifetimes of the sub-millisecond pulsars in the phase of a short
spin period }

Sub-millisecond pulsars may be very rare, or the timescale for such a
pulsar to stay in the short period phase ($<1$ ms) may not be long
enough due to magnetodipole (EM) radiation and gravitational wave
(hereafter GW) radiation (Andersson 2003).
The lowest order GW radiation is bar-mode, which is due to 
non-axisymmetric quadrupole moment.
Here we consider GW radiation on the bar mode which exerts a larger
braking torque with braking index $n\approx 5$ than magnetodipole
radiation ($n=3$).  The rotation frequency drops quickly due to GW
radiation and EM radiation:
\begin{equation}
-I\Omega \dot{\Omega} = \frac{32GI^2\varepsilon_e^2\Omega^6}{5c^5}+
\frac{B_0^2 R^6 \Omega^4}{6c^3},
\end{equation}
where $c$ is the speed of light, $\varepsilon_e=\Delta a/\bar{a}$ is
the gravitational ellipticity (equatorial ellipticity), $\Delta a$ is
the difference in equatorial radii and $\bar{a}$ is the mean
equatorial radii.

To simplify Eq. (20), we introduce the notation
$A=32GI\varepsilon_e^2/(5c^5)$ and $D=B_0^2 R^6/(6Ic^3)$, and
integrate the equation in the angular velocity's domain
$[\Omega_i=2\pi/P_{\rm i}, \Omega_0=2\pi/0.001]$, then
\begin{equation}
\tau=\frac{1}{2D}(\frac{1}{\Omega_0^2}-\frac{1}{\Omega_i^2})-\frac{A}
{2D^2}\ln{\frac{\frac{1}{\Omega_0^2}+\frac{A}{D}}{{\frac{1}{\Omega_i^2}+
\frac{A}{D}}}}.
\end{equation}

An accurate ellipticity of quark stars is unfortunately
uncertain. Nevertheless, let's estimate the $\varepsilon_e$ to
calculate the timescales in the sub-millisecond period phase for GW
and EM radiations. Cutler \& Thorne (2002) suggested $\varepsilon_e =
(I-I_0)/I_0\leq10^{-6}$. Regimbau \& de Freitas Pacheco (2003) found
from their simulations that $\varepsilon_e= 10^{-6}$ is the critical
value to have an at least one detection with interferometers of the
first generation (LIGO or VIRGO).
It was shown that direct upper limit was $\varepsilon_{\rm
  e}\simeq 1.8\times10^{-4}$ on GW emission from the Crab pulsar using
data from the first 9 months of the fifth science run of LIGO (Abbott
et al. 2008).  In addition, Owen (2005) showed that the maximum
ellipticity of solid quark stars was $\varepsilon_{\rm e,
  max}=6\times10^{-4}$. 
From the on-line catalogue hosted by the ATNF\,\footnote{{\tt
    http://www.  atnf.csiro.au/research/pulsar\\/catalogue/}}, the
seventh fastest rotating millisecond pulsar is PSR J0034-0534, which
has very low period derivative $\dot{P}\sim 4.96\times10^{-21}\, \rm
s\,s^{-1}$.  We thus use such a low $\dot{P}$ and Eq. (20) to
constrain the lower limit of the sub-millisecond pulsars' ellipticity,
which is $\varepsilon_{\rm e,min}\sim 10^{-9}$ if the stellar mass is
one order of one Solar mass.
For quark stars, in order to facilitate to compare with the neutron
stars' lifetime ($\tau$) in the phase of sub-millisecond period, we
use mean equatorial ellipticities $\varepsilon_e= 10^{-6}$ and
$\varepsilon_e= 10^{-9}$ to calculate $\tau$ for both quark stars and
neutron stars through Eq. (21). 

In the case of $\varepsilon_e= 10^{-6}$, if we make the hypothesis that
the rotational energy is lost because of EM radiation, then one can
easily derive $\tau_{\rm EM}=1/(2D)(1 /\Omega_0^2-1/\Omega_i^2) \sim
5.9\times 10^9$ yr for a typical compact star with $B_0 \sim 10^8$ G
and $M=M_\odot$. While, if we suppose that the rotational energy is
lost due to GW radiation, then $\tau_{\rm
  GW}=1/(4A)(1/\Omega_0^4-1/\Omega_i^4) \sim 10^2$ yr for a typical
compact star. The energy loss rate of GW \& EM radiation in the phase
of sub-millisecond period, for a typical compact star which has a low
magnetic field (10$^8$--10$^9$ G) either from AIC (Xu 2005) or spun
up, are $\dot{E}_{\rm GW}=32GI^2\varepsilon_e^2
\Omega^6/(5c^5)=7.0\times 10^{41}P_{\rm ms}^{-6} \rm \,erg \,s^{-1}$
and $\dot{E}_{\rm EM}= B_0^2R^6\Omega^4/(6c^3)=9.6\times10^{34}P_{\rm
  ms}^{-4}B_8^2 R_6^6 \rm \,erg\,s^{-1}$, respectively. Even if a
quark star with 1 $M_\odot$ formed from WD's AIC has a high magnetic
field such as $10^{12}$ G, the lifetime $\tau$ in the phase of
sub-millisecond is 37 years, in comparison with $\tau=336$ years for
$B_0=10^8$ G. Then the EM energy loss is similar to the GW energy loss
and becomes very important for $B_0=10^{12}$~G. For $B_0$ ranges from
$10^8$ G to $10^{11}$ G, one always has $\dot{E}_{\rm GW} \gg
\dot{E}_{\rm EM}$ for compact stars with short spin
periods ($<1$~ms). Therefore, in the case of larger
ellipticity (e. g., $\varepsilon_e= 10^{-6}$), it is clear that GW
radiation dominates the energy loss in the phase of short period for
either recycled or AIC's compact stars with low magnetic field. The
corresponding lifetime is shorter for a compact star with higher mass
($\sim M_\odot$), but longer for a star with lower mass ($\sim
0.001M_\odot$). However, if the ellipticity is lower, such as
$\varepsilon_e=10^{-9}$, EM radiation dominates the rotational energy
loss. The corresponding lifetime of a quark star (even with a high
mass $\sim M_\odot$) is long enough for us to detect.
Figure 2 shows the relation of lifetime (in the phase of $<1$\,ms) and
gravitation ellipticity $\varepsilon_{\rm e}$ for quark stars. 

In the super-Keplerian case, the timescales in the phase of $<$0.5\,ms
for quark stars with different mass are also calculated, and listed in
Table 1 (See $\tau_2$). For a high-mass quark star with larger
ellipticity, the timescale is too small for real detection; but the
timescale is $> 10^4 ~\rm yr$ for a low mass quark star. Therefore,
low mass quark stars with $\varepsilon_e \sim 10^{-6}$ could have much
longer lifetime in the phase of $<0.5$\,ms. However, for lower
ellipticity, their lifetimes in the phase of $<0.5$\,ms are long
enough for quark stars with $\sim 1M_\odot$.  Once a pulsar with spin
period $<0.5$\,ms is ever found, low mass quark stars will be
physically identified.

%------------------------------Fig 2------------------------------------
\begin{figure}
\vspace{8.0cm}
\includegraphics{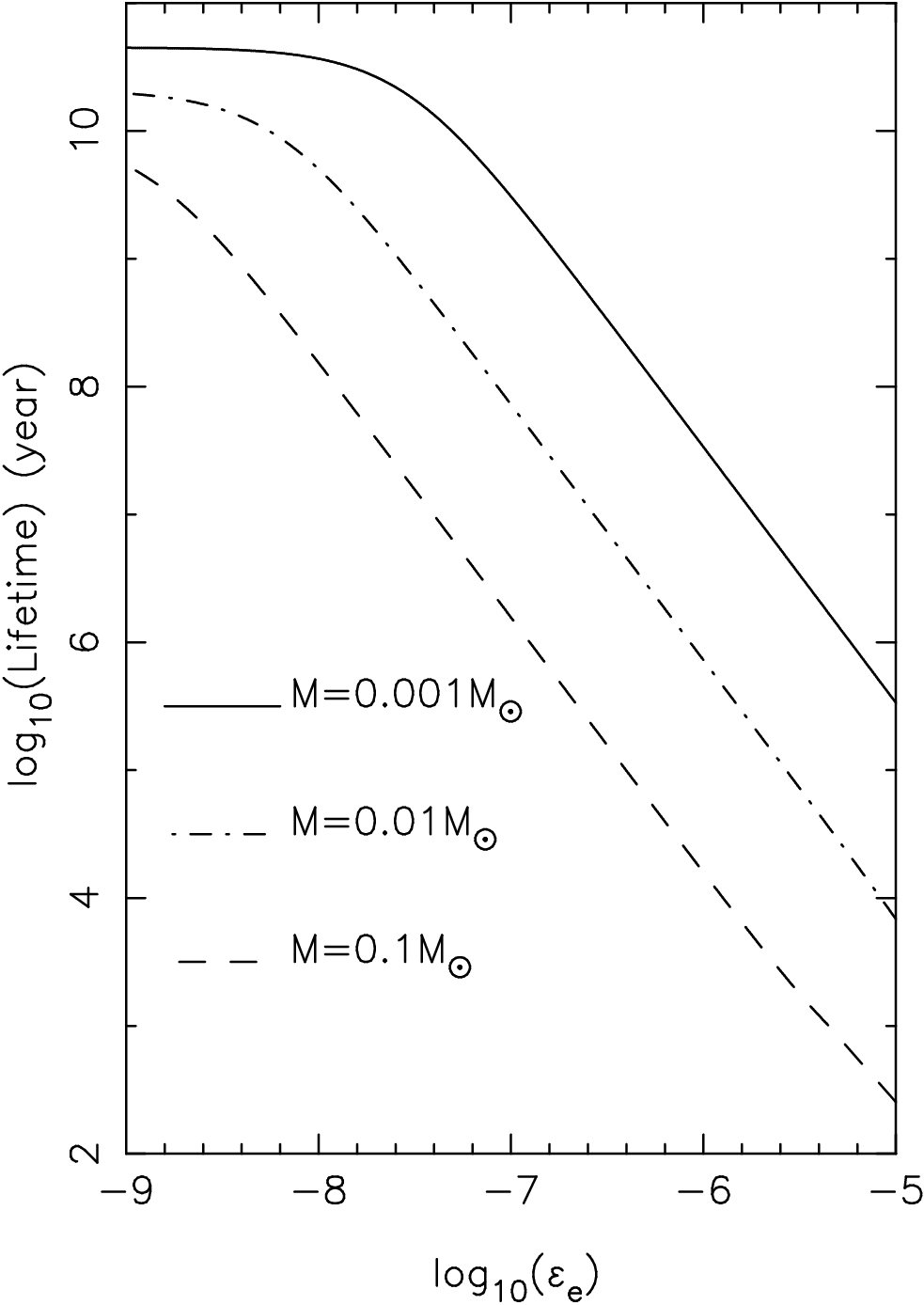}
\caption{The relation of lifetime (in the phase of $<1$\,ms) and
  gravitational ellipticity $\varepsilon_e$ for quark stars with
  masses of 0.001$M_\odot$ (solid line), 0.01$M_\odot$ (dot-dash
  line), 0.1$M_\odot$ (dashed line), magnetic field $B=10^8$G and the
  bag constant $\beta=60 \rm MeV ~fm^{-3}$. The lifetime in the phase
  of sub-millisecond period is shorter if the quark star's mass is
  higher.  } \label{fig2}
\end{figure}
%----------------------------------------------------------------

\subsection{Spin-down rate $\dot{P}$ for newborn quark stars and 
neutron stars}

%---------------------------------Fig 3-----------------------------------
\begin{figure}
\vspace{14.3cm}
\includegraphics{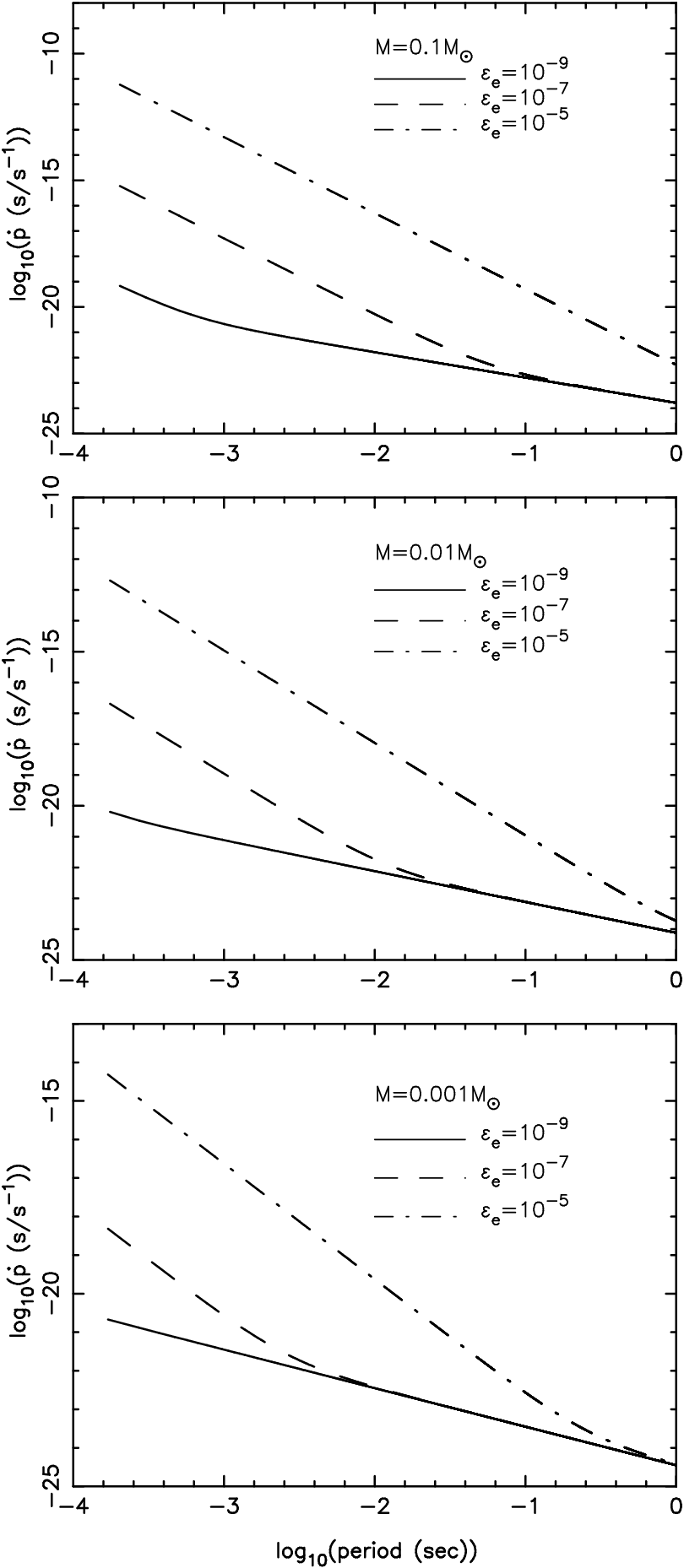}
\caption {Spin-down evolution of quark stars due to GW and EM
  radiations (period derivative versus spin period), with masses of
  0.1$M_{\odot}$, 0.01$M_{\odot}$, 0.001$M_{\odot}$. We choose
  ellipticity to be $10^{-5}$ (dot-dash lines), $10^{-7}$ (dashed
  lines), $10^{-9}$ (solid lines) in the calculation. It is evident
  that GW radiation dominates for quark stars with higher
  $\varepsilon_e$, while EM radiation dominates for lower
  $\varepsilon_e$. } \label{fig3}
\end{figure}

%----------------------------------Fig 4------------------------------------
\begin{figure}
\vspace{8.0cm}
\includegraphics{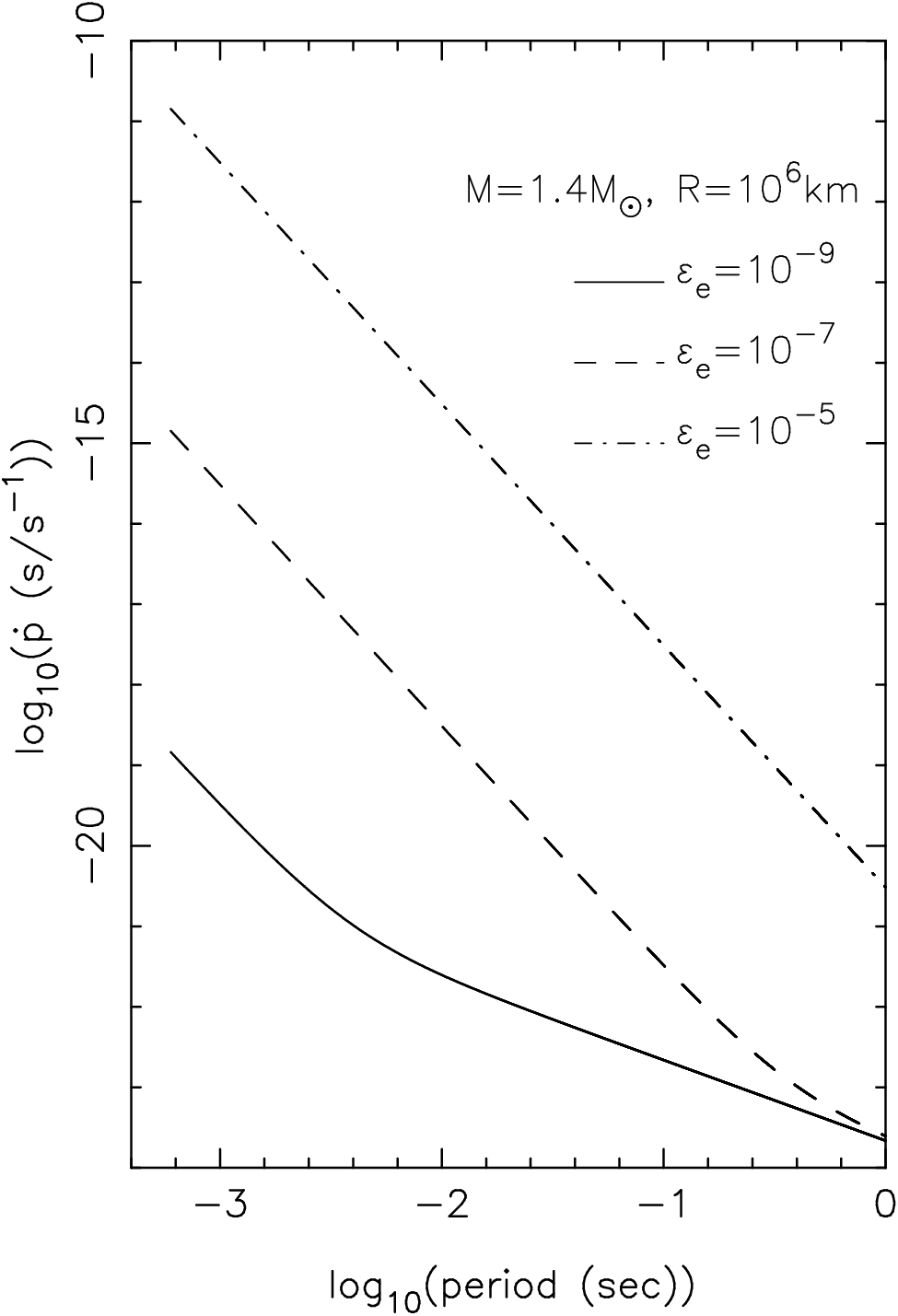}
\caption{Period derivative versus spin period diagram for a neutron
  star with an initial period of 0.5ms, mass of 1.4$M_\odot$ and
  radius of $10^6$km. The neutron star spins down quickly due to high
  mass (moment of inertia) for GW radiation.} \label{fig4}
\end{figure}
%---------------------------------------------------------------------

%{\color{blue} We also use Eq.[20] to calculate the period derivative
%(pdot) for the nascent sub-millisecond quark stars and neutron stars,
%respectively. Newborn quark stars with low mass may spin down much
%more slowly than neutron stars due to different efficiencies of GW
%radiation. From Figure 3, we can see that the spin-down rates for
%quark stars are larger if the stellar mass is higher. The dot-dash lines
%and dashed lines (with higher slope) show that GW radiation dominates
%if the $\varepsilon_e$ is higher; while EM radiation domintes for
%lower $\varepsilon_e$.
%
%As a comparison, Figure 4 is a diagram for a neutron star with an
%initial period 0.5 ms, mass of 1.4$M_\odot$ and radius of $10^6$ km,
%which varies about ten orders of magnitude. It is found that the
%neutron star spins down much more quickly than low mass quark stars,
%because of neutron star's high mass ($\sim M_\odot$) for higher
%efficiency of GW radiation.}

We also use Eq. (20) to calculate the period derivative
($\dot{P}$) for the nascent sub-millisecond quark stars and neutron
stars.

Figure 3 is a $\dot{P}-P$ diagram that shows the spin-down evolution
for quark stars with different masses. It is found that, for different
ellipticity there are different properties. For high ellipticity such
as $\varepsilon_e=10^{-5}$, the $\dot{P}$ can be changing about ten
orders of magnitude for different periods (see the steep slopes of
dot-dash lines and dashed lines). The rotational energy losses in this
case are dominated by the gravitational wave (GW) radiation. For low
ellipticity such as $\varepsilon_e=10^{-9}$, in most cases, the
rotational energy losses are dominated by magnetic dipole (EM)
radiation and the $\dot{P}$ changes with periods relatively slow (solid
lines).

 As a comparison, we also calculate the period derivative ($\dot{P}$)
 of a neutron star (with an initial period $0.5$ ms, mass of
 1.4$M_\odot$ and radius of $10^6$\,km). The results are shown in
 Figure 4. One can see that the $\dot{P}$ is changing with periods as
 large as ten orders of magnitude. It is found that the neutron star
 spins down much more quickly than low mass quark stars, because of
 neutron star's high mass ($\sim M_\odot$) for higher efficiency of GW
 radiation.

\section{Sub-millisecond pulsars formed through accretion
in binary systems}
%--------------------------------Table 3---------------------------------
\begin{table*}
 \centering
 \begin{minipage}{150mm}
 \caption
{The minimal equilibrium period for quark stars and lifetimes due to
GW and EM radiation in the phase of sub-millisecond period for quark
stars with different masses ($10^{-3}M_\odot$, $0.1M_\odot$,
$1.4M_\odot$) in the sub-Keplerian case. $\tau_1$, $\tau_2$, $\tau_3$
are calculated by using $\varepsilon_e = 10^{-6}$, while
$\tilde{\tau_1}$, $\tilde{\tau_2}$, $\tilde{\tau_3}$ are calculated by
using $\varepsilon_e = 10^{-9}$. The bag constant $\beta$ is in unit of 
$\rm Mev\,fm^{-3}$, the accretion ratio $\alpha$ is in unit of the 
Eddington accretion rate $\dot{M}_{\rm Edd}$.
\label{tbl-3}}
\begin{tabular}{cccccccccc}
\hline
$\beta$ & $\alpha$ &
$B_0(10^8{\rm G})$ & {$P_{\rm eqmin}(\rm ms)$} &
\multicolumn{1}{c}{$\tau_1(\rm yr)$} &\multicolumn{1}{c}{$\tilde{\tau_1}$(yr)}&
\multicolumn{1}{c}{$\tau_2(\rm yr)$} &\multicolumn{1}{c}{$\tilde{\tau_2}$(yr)}&
\multicolumn{1}{c}{$\tau_3(\rm yr)$} &\multicolumn{1}{c}{$\tilde{\tau_3}$(yr)}
\\
\hline
60 & 0.71 & 1.1 & 0.613 & $2.9\times10^7$ & $2.8\times10^{10}$ 
& $1.3\times10^4$ & $4.1\times10^9$ & 
$1.7\times10^2$ & $1.5\times10^8$   \\
110 & 0.85 & 1.4 & 0.453 & $5.1\times10^7$ & $3.6\times10^{10}$ 
& $2.3\times10^4$ &  $5.6\times10^9$ &
$2.3\times10^2$ & $2.6\times10^8$      \\
\hline
\end{tabular}
\end{minipage}
\end{table*}

%------------------------------Table 4-------------------------------------
\begin{table*}
 \centering
\begin{minipage}{150mm}
 \caption{The minimal equilibrium period and lifetimes for GW and
   EM radiation in the phase of sub-millisecond period of different
   EOSs of normal neutron stars in the sub-Keplerian case. The mass
   and radius data of neutron stars are obtained from Figure 2 of
   Lattimer et al. (2004). $\tau$ and $\tilde{\tau}$ are lifetimes
   within sub-millisecond period for neutron stars using
   $\varepsilon_e = 10^{-6}$ and $\varepsilon_e = 10^{-9}$
   respectively.\label{tbl-4}}
\begin{tabular}{cccccccc}
\hline 
EOS & $P_{\rm eqmin}$(ms) & Mass$(M_{\odot})$ & Radius(km) &
B$_0(10^8 \rm G)$ & $\dot{M}$($10^{17}\rm \,g\,s^{-1}$) & $\tau$(yr) & 
$\tilde{\tau}$(yr)\\
\hline
$\rm AP4$ & 0.55 & 2.21 & 10 & 2 & 6.36 & 103 & $1.02\times10^8$ \\
$\rm GS1$ & 0.52 & 1.38 & 8.27 & 2 & 5 & 255 & $2.37\times10^8$ \\
$\rm PAL6$ & 0.60 & 1.48 & 9.24 & 2 & 6.38 & 177 & $1.65\times10^8$ \\
$\rm MS0$ & 0.76 & 2.76 & 13.31 & 1 & 2.91 & 35 & $3.48\times10^7$ \\
$\rm GM3$ & 0.75 & 1.56 & 10.93 & 2 & 9.47 & 94 & $8.54\times10^7$ \\
$\rm MS1$ & 0.76 & 1.81 & 11.67 & 1 & 2.58 & 90 & $6.83\times10^7$ \\
\hline
\end{tabular}
\end{minipage}
\end{table*}

There is also an important mechanism of ``spin-up in binaries'' for
sub-millisecond pulsars' formation, which is widely discussed in the
literatures. We regard this as ``sub-Keplerian case'' and
make a comparison with our proposed AIC model ``super-Keplerian
case''. In this section, we will find the minimal periods for both
neutron stars and bare quark stars spun up by accretion in binary
systems.  We assume that the initial rotational periods of newborn
pulsars could have an ``equilibrium period'' with
two characteristic parameters: magnetospheric radius and corotation
radius. The magnetospheric radius $(r_m)$ is the radius where the ram
pressure of particles is equal to the local magnetic pressure, i.e.
\begin{eqnarray}
\nonumber r_m &=& \phi R_A = \phi(\frac{4\mu_m^2
M^{3/2}}{\dot{M}\sqrt{2G}})^{2/7}
= \phi (\frac{B_0^2R^6}{\dot{M}\sqrt{2GM}})^{2/7}\\
& = & \Big\{ \begin{array}{l} 3.24\times 10^8\phi B_{12}^{4/7}
M_1^{-1/7}R_6^{12/7}
\dot{M}_{17}^{-2/7} ~\rm cm,  \\
 1.857\times10^6\phi B_8^{4/7} M_1^{3/7} \beta_{14} ^
{-4/7}\dot{M}_{17}^{-2/7} ~\rm cm,
\end{array}
\end{eqnarray}
where $\mu_m$ is the magnetic moment of per unit mass of the compact
star; $B_8$ is the surface magnetic strength in units of $10^8$ G and
$\dot{M}_{17}$ is the accretion rate in units of $10^{17} ~\rm
g~s^{-1}$; $\phi$ is the ratio between the magnetospheric radius and
the Alfv$\rm \acute{e}$n radius (Wang 1997; Burderi \& King
1998). Wang (1997) studied the torque exerted on an oblique
rotator and pointed out that $\phi$ decreased from 1.35 to 0.65 as the
inclination angle increased from $0^{\circ}$ to $90^{\circ}$.  Here we
take $\phi \sim 1$, the influence of $\phi$ is discussed in $\S 6$.

When $r_m$ is very close to the compact star's radius, we could
rewrite the accretion rate $\dot{M}$ in units of Eddington accretion
rate ($\dot{M}_{\rm Edd}$), with a ratio, $\alpha$, so that
\begin{equation}
\dot{M} = \alpha \dot{M}_{\rm Edd} = \alpha \frac{4\pi c
m_pR}{\sigma_T} = 1.0\times 10^{18}\alpha M_1^{1/3} \beta^ {-1/3}
~\rm g~s^{-1}.
\end{equation}
With these equations obtained above, then we can get $r_m$ for quark
stars,
\begin{equation}
r_m = 9.6\alpha^{-2/7}B_8^{4/7} M_1^{1/3} \beta_{14}^{-10/21} ~\rm
km.
\end{equation}
The corotation radius is $r_c = 1.5\times10^8M_1^{1/3} P^{2/3}$ cm.
The spin periods of compact stars cannot exceed the Kepler
limit via accretion. When the compact star was spun up to the Kepler
limit by the accreted matter falling onto the compact star's
surface, for neutron stars, as the equatorial radius expanded, one
can use the simple empirical relation for the maximum spin frequency

\begin{equation}
\Omega_{\rm max}=7700M_1^{1/2}R_6^{-3/2} ~ \rm s^{-1}
\end{equation}
(Haensel \& Zdunik 1989; Lattimer \& Prakash 2004), which leads to
\begin{equation}
P_{\rm eq} \geqslant 0.816 M_1^{-1/2}R_6^{3/2} ~\rm ms,
\end{equation}
where $M$ and $R$ refer to the neutron star's mass and radius of
nonrotating configurations.

For quark stars, Gourgoulhon et al. (1999) used a highly precise
numerical code for the 2-D calculations, and found that the
$\Omega_{\rm max}$ could be expressed as
$\Omega_{\rm max}=9920\sqrt{\beta_{60}} ~\rm rad~s^{-1}$, where
$\beta_{60}=\beta/(60 ~{\rm MeV~fm^{-3}})$, which implied that
$P_{\rm eq} \geqslant 0.633{\beta_{14}^{-1/2}}~\rm ms$. These are the
so-called ``sub-Keplerian condition''.

The accretion torque, $N$, exerted on the compact star contains two
contributions: one is positive material torque which is carried by
the materials falling onto the star's surface; the other is magnetic
torque which can be positive or negative, depending on the fastness
parameter $\omega_s=\Omega_\star/\Omega_{\rm K}=(r_m/r_c)^{3/2}$. It
is suggested that all the torques may cancel one another if the fastness
is $\omega_s=(r_m/r_c)^{3/2}\approx 0.884$ (Dai \& Li 2006). This
implies a magnetospheric radius of $r_m = 0.92 r_c \approx r_c$. One
can obtain an equilibrium period of $P_{\rm eq}$ when setting $r_m =
r_c$,
\begin{eqnarray}
P_{\rm eq} = \Big\{ \begin{array}{lr}
0.512B_8^{6/7}\beta_{14}^{-5/7}\alpha^{-3/7}
~\rm ms, &(a)\\
3170B_{12}^{6/7}M_1^{-5/7} R_6^{18/7}\dot{M}_{17}^{-3/7} ~\rm ms.  &
(b)
\end{array}
\end{eqnarray}
For quark stars, the equilibrium period is independent of mass and
radius, and only dependent on bag constant, surface magnetic field,
and accretion rate. Take $B_0$ in the range $[10^8 ~\rm G,
10^{12} ~\rm G]$, we may use Eq. (27a) to calculate the minimal
equilibrium period of different EOSs (equation of state) for quark
stars. For $\beta=60 ~\rm MeV~fm^{-3}$, when $\alpha = 0.71, B_8 =
1.1$, one can get the minimal period 0.613 ms. For $\beta=110 ~\rm
MeV~fm^{-3}$, when $\alpha = 0.85, B_8 = 1.4$, the minimal period is
0.453 ms. (See results in Table 3.)

For neutron stars, data for mass and radius in different EOSs were
taken from Lattimer \& Prakash (2004, their Figure 2), $B_0$ is in the
range $[10^8 ~\rm G, 10^{12} ~\rm G]$.  The minimal equilibrium
period is calculated using the Eq. (27b). (See results in Table 4.)

In the sub-Keplerian case, the timescales in the phase of
sub-millisecond for quark stars of different mass and neutron stars of
different EOSs are listed in Table 3 and Table 4, respectively. 
For typical quark stars as well as neutron stars with
high $\varepsilon_e$, their lifetimes in the phase of sub-millisecond
period are about $10^2$ years, which result in a too low detection
possibility. However, for low $\varepsilon_e$, the lifetime of a 
sub-millisecond pulsar (even with a high mass) is long enough.

\section{Conclusions and Discussions}

%Sub-millisecond pulsars are probably quark stars, distinguishable
%from neutron stars. 
If a sub-millisecond pulsar is ever found, we have shown that it
could be a quark star based upon plausible scenarios for its origin,
the energy available for radiation and its lifetime.  A new possible
way to form sub-millisecond pulsars (quark stars) via AIC of white
dwarfs has been discussed in this paper. In the super-Keplerian case,
we derived the initial period $P_{\rm q}$ via angular momentum
conservation with consideration of the special and general
relativistic effects, and calculated the lifetime and gap parameters
of a newborn quark star.  Quark stars with different masses could have
the minimal rotational period around 0.1 ms. In most cases, quark
stars would be bare (Xu 2002), therefore, a vacuum gap would be formed
in the polar cap region. Based on our rough estimations without
considering the effect of frame dragging (Harding \& Muslimov 1998),
we found that the basic parameters (including rotational energy loss)
in the gap are suitable for pair (electrons and positrons) production
and sparking. They can be detected as sub-millisecond radio pulsars.

We also used an approximate formula to calculate nascent quark star's
moment of inertia, but there are no accurate solutions to fast
rotating compact stars' configuration until nowadays. It should be
investigated precisely in the future. In the calculation of WD's mass
and radius, we just considered the non-rotating configuration. But it
does not change the conclusions of this paper.  If the central density
$\rho_{\rm c}$ of the WD is lower than $10^{11}$~g~cm$^{-3}$ before
collapsing, the resulting WD has a larger radius and moment of
inertia, consequently, the newborn quark star could have a smaller
spin period ($<$1\,ms).

Both the special and general relativistic effects are weak for a low
mass (e.g. Jupiter-like) quark star with a small radius. The
rotational energy is lost via GW and EM radiation. The GW radiation
dominates the rotational energy loss in the phase of sub-millisecond
period, if magnetic field of stars is not so large. Such quark stars
therefore have long lifetimes (several million years if mass $\sim
10^{-3}M_\odot$) to maintain their spin periods of sub-millisecond. We
have considered the bar-mode of GW radiation in this paper, while
other GW mode (e.g. r-mode) may be important but not yet considered
here ~\citep{xu06gw}. The subsequent relaxation timescale of a newborn
quark star to a rigidly rotating configuration could be negligible
since a quark star may be solidified soon after birth.

An important constraint for sub-millisecond pulsar's detection is
its lifetime in the phase of $<1$\,ms due to GW and EM radiation. A
possible method is proposed to constrain the lower limit of the
pulsars' equatorial ellipticity, i. e., $\varepsilon_{\rm e,min}\sim
10^{-9}$, by evaluating millisecond pulsars' period derivative via
Eq. (20).  For larger ellipticity, e. g., $\varepsilon_e= 10^{-6}$, it
is clear that GW radiation dominates the energy loss in the phase of
short period for either recycled or AIC's compact stars.  The
corresponding lifetime is shorter for a compact star with higher mass
($\sim M_\odot$), but longer for a star with lower mass ($\sim
0.001M_\odot$). However, if the ellipticity is lower, e. g.,
$\varepsilon_e=10^{-9}$, EM radiation dominates the rotational energy
loss. The corresponding lifetime of a quark star (even with a high
mass $\sim M_\odot$) is long enough, and there are no lifetime
constraints for sub-millisecond pulsars' detection. Solid evidence of
quark stars will be obtained if a pulsar with a period of less than
$\sim 0.5$ ms is discovered in the future.

In the sub-Keplerian case, neutron and ``bare'' quark stars can be
spun up to sub-millisecond periods (even $\sim 0.5$ ms) through
accretion in binary systems. When neutron stars are spun up to the
Kepler limit, the minimal equilibrium periods depend only on the mass
and radius of the nonrotating configurations. Quark stars' minimal
equilibrium periods depend on the bag constant.

\section*{Acknowledgments}
The authors are very grateful to the referee for valuable comments.
We thank for useful conversations at both the pulsar groups of NAOC
and of Peking University. We are also grateful to Prof. Gao, C. S. for
a valuable discussion. Especially, we appreciate Prof. Chou, Chih Kang 
for improving our language. This work is supported by NSFC (10521001,
10573002, 10778611, 10773016 and 10833003) and the Key Grant Project
of Chinese Ministry of Education (305001).
% and the program of the Light in China's Western Region (LCWR, No.
%LHXZ200602).

\appendix
\section{Differentially Rotating WD Model}

WD could be rotating differentially. As stated by (Mueller \& Eriguchi
1985), the WD's angular velocity $\Omega$ is a function of the
distance from the rotation axis $\widetilde{\omega}$. The angular
momentum distribution (so-called rotation law) is
\begin{equation}
\Omega(\widetilde{r})=\Omega_{\rm c}\frac{(a R_{\rm e
})^2}{(a R_{\rm e})^2+\widetilde{r}^2},
\end{equation}
where $\Omega_c$ is the central angular velocity, $R_e$ is the
equatorial radius, and $a$ is a free parameter. When differential
rotation is taking into account, we can numerically evaluate the
angular momentum of the WD's inner collapsed core, i.e.,
\begin{eqnarray}
&J_{\rm core}&=\sum_{\rm i} J_{\rm i}=\sum_{\rm i} \int_0 ^{\pi}
  {\sigma2\pi r^4_i\sin^3{\theta}\Omega(r_i\sin{\theta})}\rm
  d\theta\\\nonumber &&=\sum_{\rm i}[ \frac{m_{\rm core}\Omega_c a^2
      R_{\rm WD}^2}{r_i^2}\times\\\nonumber & & (r^2_i - 0.5
    \sqrt{\frac{r^2_i}{a^2R_{\rm WD}^2+r^2_i} }a^2R_{\rm
      WD}^2\ln{\frac{1+\sqrt{\frac{r^2_i} {a^2R_{\rm WD}^2+r^2_i}}
      }{1-\sqrt{\frac{r^2_i} {a^2R_{\rm WD}^2+r^2_i}} } } )],
\end{eqnarray}
where $J_i$ is the angular momentum of each spherical shell with
integral radius $r_i$. 

\label{lastpage}

\end {document}